\documentclass{PoS}
\usepackage{amssymb}

\title{The effective $U(1)$-Higgs theory at strong coupling on optical lattices?}

\ShortTitle{The effective $U(1)$-Higgs theory at strong coupling on optical lattices?}

\author{\speaker{Alexei Bazavov}$^{a,b}$, Chen-Yen Lai$^a$, Shan-Wen Tsai$^a$, Yannick Meurice$^b$\\
\llap{$^a$}Department of Physics and Astronomy, University of California, Riverside, CA 92521, USA\\
\llap{$^b$}Department of Physics and Astronomy, The University of Iowa, Iowa City, Iowa 52242, USA\\
        E-mail: \email{obazavov@quark.phy.bnl.gov}
}

\abstract{
We discuss the $U(1)$-Higgs model in two dimensions in the strongly coupled regime.
If we neglect the plaquette interactions, we generate an effective theory where link variables are integrated out,
producing 4-field operators. Plaquette interactions can be restored order by order as in recent calculations with staggered fermions. In the case of a $SU(2)$ gauge theory with fermions, this strong coupling expansion can be related to the strong coupling expansion of Fermi-Hubbard models possibly implementable on optical lattice. We would like to provide a similar construction relating the $U(1)$-Higgs model to some Bose-Hubbard model. As a first step in this direction, we discuss a recent proposal to implement the $O(2)$ model on optical lattices using a $^{87}$Rb and $^{41}$K
Bose-Bose mixture of cold atoms.}

\FullConference{The 32nd International Symposium on Lattice Field Theory\\
		 23-28 June, 2014\\
		 Columbia University New York, NY}

\begin{document}

\section{Introduction}

Recent advances in the technology of engineering many-body systems with
cold atoms trapped in optical lattices allow for building quantum
simulators, \textit{i.e.}, systems with model quantum Hamiltonians, where
types of interactions can be customized and their strengths
tuned.
If systems, whose Hamiltonians resemble or approximate well, for example,
condensed matter or lattice gauge theory models, can be realized experimentally,
they would serve as ``analog computers'', providing answers by quantum
mechanical measurements, rather than conventional perturbative or Monte Carlo
techniques (routinely being done nowadays on digital computers).
Recent proposals along these lines are discussed in 
Refs.~~\cite{Zou:2014rha,banerjee2013,zohar2013,tagliacozzo2013,Liu:2012dz}.

For lattice gauge theory, in particular, such program of building quantum
simulators requires several steps. First, most lattice calculations utilize
the path integral quantization in the imaginary-time Lagrangian formulation,
while cold atom systems in a lab evolve according to their quantum Hamiltonians in
real time. Therefore, one needs to revive the Hamiltonian approach to lattice
gauge theory, pioneered by Kogut and Susskind in~\cite{Kogut:1974ag}.
Second, cold atoms in optical lattices reside in a periodic potential, similar to
the one that electrons feel in a crystalline solid. Thus, one expects that models used in
condensed matter, \textit{e.g.}, the Hubbard model, would be the closest ones
to typical model Hamiltonians that can be realized on optical lattices. For this
reason one needs to find suitable mappings from lattice gauge theory models to
condensed matter-like ones. A gateway to this is an observation~\cite{PhysRevB.38.2926}
that the Fermi-Hubbard model and $SU(2)$ lattice gauge theory share the same
strong-coupling expansion. And last, but not least, the model Hamiltonians
should be simple enough to be feasible for experimental realization.

\section{Gauge-Higgs models}

Let us consider the $U(1)$-Higgs model with the action:
\begin{eqnarray}
S &=& -\beta\sum_x\sum_{\nu<\mu}{\rm Re}{\rm Tr}\left[U_{x,\mu\nu}\right]
+ \lambda\sum_x\left(\phi_x^\dagger\phi_x-1\right)^2+
\sum_x\phi_x^\dagger\phi_x\nonumber\\
&-&\kappa\sum_x\sum_{\mu=1}^{d}
\left[\phi_x^\dagger U_{x,\mu}\phi_{x+\hat\mu}+
\phi_{x+\hat\mu}^\dagger U^\dagger_{x,\mu}\phi_x
\right].
\end{eqnarray}
The path integral quantization is then
\begin{equation}
Z=\int D\phi^\dagger D\phi DU e^{-S}.
\end{equation}

At the lowest order of the strong-coupling expansion we set $\beta=0$ and carry out
$DU$ integration. For a particular link $U_{x,\mu}$ we have an integral:
\begin{equation}
J\equiv \int dU \exp\left\{\kappa\left(\phi^\dagger U\psi +
\psi^\dagger U\phi\right)\right\},
\end{equation}
where for simplicity we define $U\equiv U_{x,\mu}$, $\phi\equiv\phi_x$ and
$\psi\equiv\phi_{x+\hat\mu}$. The measure is such that $\int dU=1$ and
$\int UdU=0$.

If we expand the exponent, only terms that have same power of $U$ and $U^\dagger$
produce a non-zero contribution. The expansion is:
\begin{equation}
J=\sum_{n=0}^{\infty}\frac{1}{(n!)^2}\left(\kappa^2\phi^\dagger\phi\psi^\dagger\psi\right)^n
=I_0\left(2\kappa\sqrt{\phi^\dagger\phi\psi^\dagger\psi}\right).
\end{equation}

Thus, after integrating the gauge field we have the following partition function:
\begin{equation}
Z = \int D\phi^\dagger D\phi \exp\left\{-\lambda\sum_x (\phi_x^\dagger\phi_x-1)^2
-\sum_x\phi_x^\dagger\phi_x+\sum_x\sum_{\mu=1}^{d}\ln
I_0\left(2\kappa\sqrt{\phi_x^\dagger\phi_x\phi_{x+\hat\mu}^\dagger\phi_{x+\hat\mu}}\right)\right\}.
\end{equation}
At small $\kappa$ we can keep the first non-trivial order only and
we have at $O(\kappa^4)$:
\begin{equation}
Z_{EFT}=\int D\phi^\dagger D\phi \exp\left\{-\sum_x[\lambda(\phi_x^\dagger\phi_x-1)^2
+\phi_x^\dagger\phi_x-\kappa^2
\sum_{\mu=1}^{d}\phi_x^\dagger\phi_x\phi_{x+\hat\mu}^\dagger\phi_{x+\hat\mu}]\right\}.
\end{equation}
If we write $\phi_x=|\phi_x|{\rm e}^{i\theta_x}$, we see that the Nambu-Goldstone modes $\theta_x$ have completely disappeared from the $\beta=0$ effective action which depends only on the ``meson'' operator $M_x=\phi_x^\dagger \phi_x$. The integration over the gauge fields generates powers of   $M_x M_{x+\hat \mu}$ in the effective action. 

$M_x M_{x+\hat \mu}$ terms are also generated in the same approximation for  $SU(N)$ gauge theories with {\it fermions}. In that case, the meson operator is  $M_x=\bar{\psi_x}\psi_x$ but in addition we have baryon-baryon interactions $B_x B^{\dagger}_{x+\hat \mu}$with the baryon operator $B_x =\epsilon_{i_1i_2\dots i_N}\psi_x^{i_1}\psi_x^{i_2}\dots \psi_x^{i_N} $~\cite{Rossi:1984cv}.

It has been pointed out \cite{PhysRevB.38.2926} that the Heisenberg Hamiltonian, which appears at second order in degenerate perturbation theory of the 
Fermi-Hubbard model with strong onsite repulsion:
\begin{equation}
H = J\sum_{<\mathbf{i}\mathbf{j}>} \mathbf{S}_\mathbf{i}\cdot \mathbf{S}_\mathbf{j} \  \  \  \  {\rm with}\    \   J=4t^2/U
\end{equation}
can also be derived in the strong coupling limit of a $SU(2)$ lattice gauge theory. 
Using $\mathbf{S}_\mathbf{i} = \frac{1}{2} f_{\mathbf{i}\alpha}^\dagger \mathbf{\sigma}_{\alpha \beta} f_{\mathbf{i}\beta}$, 
imposing the constraint $f_{\mathbf{i}\alpha}^\dagger f_{\mathbf{i}\alpha} =1$ and a particle-hole transformation, one obtains 

\begin{equation}
H= \frac{J}{8} \sum_{x,\hat{\mu}}\left[M_x M_{x+\hat{\mu}}+2(B_x^\dagger B_{x+\hat{\mu}}+B_{x+\hat{\mu}}^\dagger B_{x})\right]-\frac{Jd}{4}\sum_x\left(M_x-\frac{1}{2}\right) \ 
\label{eq:29}
\end{equation} 

The order $\beta$ corrections to the  effective action for $SU(N)$ theories with fermions are being studied, 
{\it e.g.} Ref.~\cite{deForcrand:2014tha}.
In the Abelian case, one can use tensor renormalization group methods~\cite{Liu:2013nsa}
to calculate the first correction to the partition function due to the plaquette interaction. It has the form 
\begin{equation}
\prod_{<xy>\in pl.}I_1(2\kappa|\phi_x||\phi_y|).
\end{equation}
We now turn to a simpler model sharing some features with the $U(1)$-Higgs model to illustrate how it can be connected to models
realized on optical lattices (see Ref. \cite{Zou:2014rha}  for more details).

\section{Classical $O(2)$ model in 1+1 dimension}

The partition function of the model is
\begin{equation}
    Z = \int{\prod_{(x,t)}{\frac{d\theta_{(x,t)}}{2\pi}} {\rm e}^{-S}}\ ,
\label{eq:bessel}
\end{equation}
\begin{equation}
S = -\beta_t \sum\limits_{(x,t)} \cos(\theta_{(x,t+1)} - \theta_{(x,t)}-i\mu)
-\beta_x \sum\limits_{(x,t)} \cos(\theta_{(x+1,t)} - \theta_{(x,t)}),
\end{equation}
where $\mu$ is the chemical potential.
The sites of the rectangular $N_x\times N_t$ lattice are labeled as $(x,t)$ and we assume periodic boundary conditions in space and time. We take $\beta_t \gg \beta_x$ and obtain the time continuum limit.
To quantize the model we promote $\theta$ variables to operators and arrive at the
Hamiltonian connecting quantum rotors  on a lattice with $\beta_x$ acting as the coupling between the spatial sites: 
\begin{equation}
\hat{H}=\frac{\tilde{U}}{2}\sum_x \hat{L}_{(x)}^2-\tilde{\mu}\sum_x \hat{L}_{(x)}-\tilde{J}\sum_{\left<xy\right>}\cos(\hat{\theta}_{(x)}-\hat{\theta}_{(y)}) \ ,
\label{eq:rotor}
\end{equation}
with $\tilde{U}=1/(\beta_t a)$, $\tilde{\mu}=\mu/a$ and $\tilde{J}=\beta_x/a$, the sum extending over sites $x$ and nearest neighbors $\left<xy\right>$ of the space lattice
and $a$ is a lattice spacing.
The operator $\hat L=-i\partial/\partial\theta$ is similar to the angular momentum
operator. Its eigenstates $\hat L|m\rangle=m|m\rangle$ span an infinite dimensional
Hilbert space and $m$ takes all positive and negative integer values.

For realistic implementations with cold atoms,  it is convenient to consider
Hamiltonians with operators that live in a finite
rather than infinite Hilbert space \cite{Orland:1989st,Chandrasekharan:1996ih}.
The $O(2)$ model is the simplest, nontrivial, model where Abelian, quantum link inspired,
projections can be introduced, and we illustrate the strategy below.

The third interaction term in the Hamiltonian (\ref{eq:rotor}) can be written in
terms of $e^{\pm i\hat\theta}$ operators. They satisfy the following algebra:
\begin{equation}\label{L_comm}
[\hat L, e^{\pm i\hat\theta}]=\pm e^{\pm i\hat\theta},\,\,\,\,\,
e^{i\hat\theta}e^{-i\hat\theta}=1.
\end{equation}
From the commutation relations we find that $e^{\pm i\hat\theta}$ act as
ladder operators:
\begin{equation}
\hat Le^{\pm i\hat\theta}|m\rangle = \left(\pm e^{\pm i\hat\theta}+
e^{\pm i\hat\theta}\hat L\right)|m\rangle = (m\pm1)e^{\pm i\hat\theta}|m\rangle,
\end{equation}
\begin{equation}
e^{\pm i\hat\theta}|m\rangle = |m\pm1\rangle,\,\,\,\,\,
\langle m\pm 1|m\pm1\rangle = \langle m|e^{\mp i\hat\theta}e^{\pm i\hat\theta}|m\rangle =1,
\end{equation}
with the transition matrix elements equal to 1.

Consider now the interaction term connecting sites $x$ and $y$:
\begin{equation}
\hat C_{xy}\equiv
\cos(\hat\theta_{(x)}-\hat\theta_{(y)})=\frac{1}{2}\left\{
e^{i(\hat\theta_{(x)}-\hat\theta_{(y)})}+e^{-i(\hat\theta_{(x)}-\hat\theta_{(y)})}\right\}.
\end{equation}
The Hilbert space in this case is
\begin{equation}
|m_x,m_y\rangle = |m_x\rangle\otimes |m_y\rangle.
\end{equation}
and the matrix elements of $\hat C_{xy}$ can be easily found from
\begin{equation}\label{C12}
\hat C_{xy}|m_x,m_y\rangle = \frac{1}{2}\left(|m_x-1,m_y+1\rangle+|m_x+1,m_y-1\rangle\right).
\end{equation}
In the $|m_x,m_y\rangle$ basis the $\hat C_{xy}$ operator is an infinite matrix that has
a constant above and below the main diagonal, with all other matrix elements equal to 0.
We need to approximate this infinite matrix with a finite one. We can choose
a maximum value $m_{max}$ and truncate the matrix keeping only
$-m_{max}\leqslant m_i\leqslant m_{max}$
entries. Such an approximation converges to the original $\hat C_{xy}$ in the limit
$m_{max}\to\infty$, however, it does not fully respect the original algebra (\ref{L_comm}). This can be seen, for instance, by acting with 
$e^{i\hat\theta}e^{-i\hat\theta}$ on the state of lowest $m$: this gives 0 in contradiction with the identity $e^{i\hat\theta}e^{-i\hat\theta}=1$.

Alternatively, we can look for an approximation that slightly modifies the original algebra
with a possible expense of less accurately reproducing the matrix elements of $\hat C_{xy}$.
The structure of $\hat C_{xy}$ looks very similar to the action of the
raising and lowering operators $\hat L^\pm$ of the angular momentum algebra, which is also
similar to (\ref{L_comm}). Moreover, in that case one naturally has $l=m_{max}$.
Thus, one can argue that the truncated operator $\hat C_{xy}$ can be approximated as the one
made of $\hat L^\pm$ operators in the representation $l=m_{max}$. In this case the basis states
are the spherical harmonics $|lm\rangle$ and we replace $\hat L$ with $\hat L^z$ and
the original raising and lowering operators $e^{\pm i\hat\phi}$ with $\hat L^\pm$:
\begin{equation}\label{L_comm2}
[\hat L^z, \hat L^\pm]=\pm \hat L^\pm,\,\,\,\,\,
[\hat L^+,\hat L^-]=2\hat L^z.
\end{equation}
The matrix element of the raising operator is defined as:
\begin{equation}\label{Lpm_ME}
\langle l,m | L^+|l,m-1\rangle = \sqrt{(l+m)(l-m+1)}.
\end{equation}

Consider $l=1$ case. We have three states, $m=-1,0,1$ and all the matrix
elements are equal to $\sqrt{2}$, following from (\ref{Lpm_ME}),
and, thus, the truncated operator can be represented exactly.
(This is no longer true for higher representations.)

We embed the original basis $|m\rangle$, $|m|\leqslant1$ into the spherical harmonics basis
$|l=1,m\rangle$ (with the identification $\hat L^z\equiv\hat L$ we have the correct
eigenvalues $\hat L^z|lm\rangle=m|lm\rangle$)
and represent the original $\hat C_{xy}$ operator with:
\begin{equation}
\hat{\tilde C}_{xy}=A\left(\hat L^+_{(x)}\hat L^-_{(y)}
+\hat L^-_{(x)}\hat L^+_{(y)}\right).
\end{equation}
By acting on, for example, $|l_x=1,m_x=0;l_y=1,m_y=0\rangle$ state we can easily
deduce that $A=1/4$.

We have then the following Hamiltonian:
\begin{equation}
\hat{H}=\frac{\tilde{U}}{2}\sum_x \left(\hat{L}^z_{(x)}\right)^2
-\tilde{\mu}\sum_x \hat{L}^z_{(x)}
-\frac{\tilde{J}}{4}\sum_{\left<xy\right>}
\left(\hat L^+_{(x)}\hat L^-_{(y)}+\hat L^-_{(x)}\hat L^+_{(y)}\right) \ .
\label{eq:rotor2}
\end{equation}

\section{Two-species Bose-Hubbard model}

The Hamiltonian (\ref{eq:rotor2}) of the $O(2)$ model can be realized in optical lattice
experiments if an appropriate mapping to the Bose-Hubbard model can be found.
Interpreting the  
positive (negative) eigenvalues of $\hat L^z$  as the charges of particles  (antiparticles) states associated with a complex scalar field, it is natural to consider a two-species Bose-Hubbard Hamiltonian on a lattice. We use the following parameterization:
\begin{equation}\label{2bh}
\mathcal{H}=-\sum_{\langle xy\rangle}(t_a a^\dagger_x a_y+t_b b^\dagger_x b_y+h.c.)-\sum_{x, \alpha}(\mu + \Delta_\alpha) n^\alpha_x
+\sum_{x, \alpha}\frac{U_\alpha}{2}n^\alpha_x(n^\alpha_x-1)+W\sum_xn^a_xn^b_x+\sum_{\langle xy\rangle,\alpha}V_\alpha n^\alpha_xn^\alpha_y \,
\end{equation}
with $\alpha=a,b$ indicating the two different species, 
 $n^a_x=a^\dagger_xa_x$ and $n^b_x=b^\dagger_xb_x$ the number operators, and $|n^a_x,n^b_x\rangle$ the corresponding on-site basis. 
It is possible to adjust the chemical potentials in order to set $\langle n_x\rangle = \langle n_x^a + n_x^b\rangle = 2$. 
 In the limit where $U_a=U_b=W$ are very large and positive, 
the on-site Hilbert space can then be restricted to the states satisfying $n_x=2$ at each site. All the other states (with $n_x\ne 2$) belong to high-energy sectors that are separated from this one by energies of order $U$. The three states 
 $|2,0\rangle$, $|1,1\rangle$ and $|0,2\rangle$ 
 correspond to the three states of the spin-1 projection considered above.

Let us restrict the model to just two sites $x$ and $y$ for the moment. The Hilbert space consists
of nine states:
\begin{equation}
|n^a_x,n^b_x;n^a_y,n^b_y\rangle=|n^a_x,n^b_x\rangle\otimes|n^a_y,n^b_y\rangle,\,\,\,\,\,
\mbox{with}\,\,\,\,\,|n^a_i,n^b_i\rangle=|2,0\rangle,|1,1\rangle,|0,2\rangle,\,\,\,i=x,y.
\end{equation}
For $U,W\gg t_\alpha$ we consider hopping as perturbation, split the
Hamiltonian as:
\begin{eqnarray}
\mathcal{H}_0&=&\sum_{x, \alpha}(\mu+\Delta_\alpha) n^\alpha_x
+\frac{U_0}{2}\sum_{x,\alpha} n^\alpha_x(n^\alpha_x-1)
+W\sum_x n_x^an_x^b+\sum_{\langle xy\rangle \alpha}V_\alpha n_x^\alpha n_y^\alpha,\\
\mathcal{H}_I&=&-\sum_{\langle xy\rangle}(t_a a^\dagger_x a_y+t_b b^\dagger_x b_y+h.c.),
\end{eqnarray}
and proceed with the degenerate perturbation theory. (In the regime of strong onsite repulsion
$U_0\gg (U_0-W), V_\alpha$ this basis is approximately degenerate.)

At zeroth order in $t_\alpha$ the matrix elements of the effective Hamiltonian are
given by the action of various number operators on the basis states. At first order
the hopping term generates transitions into the states with higher occupation, that
are more energetically costly due to large $U_0$. These are ``virtual'' states
and in the chosen $n=2$ subspace the first order contribution is zero.
At the second order transitions to the virtual states and back are allowed, and,
thus, the hopping generates non-zero non-diagonal matrix element in $\mathcal{H}_I$.
Overall, the terms generated at the second order can be mapped onto $L^z$, $L^\pm$
operators. The effective second order Hamiltonian (generalized for arbitrary
number of sites and arbitrary fixed occupation number):
\begin{eqnarray}\label{H_eff}
\mathcal{H}_{eff}&=&
\left(\frac{V_a}{2}-\frac{t_a^2}{U_0}+\frac{V_b}{2}-\frac{t_b^2}{U_0}\right)
\sum_{\langle xy\rangle}L^z_{(x)}L^z_{(y)}+
\frac{-t_at_b}{U_0}\sum_{\langle xy\rangle}(L^+_{(x)}L^-_{(y)}+L^-_{(x)}L^+_{(y)})
+(U_0-W)\sum_{x}(L^z_x)^2\nonumber\\
&+&\left[\left(\frac{pn}{2}V_a+\Delta_a-\frac{p(n+1)t_a^2}{U_0}\right)
-\left(\frac{pn}{2}V_b+\Delta_b-\frac{p(n+1)t_b^2}{U_0}\right)\right]\sum_{x}L^z_{(x)},
\end{eqnarray}
where $p$ is the number of neighbors and $n$ is the occupation ($p=2$, $n=2$ in our case).
$\hat L$ is the angular momentum operator in representation $n/2$.

Notice that the effective Hamiltonian (\ref{H_eff}) is very similar to (\ref{eq:rotor2})
but contains an extra $\hat L^z\hat L^z$ term. By choosing the hopping amplitude 
$t_\alpha = \sqrt{V_\alpha U_0/2}$ it can be removed and we have
\begin{eqnarray}
\mathcal{H}_{eff}=\frac{U}{2}\sum_x(L^z_{(x)})^2-\widetilde{\mu}\sum_xL^z_{(x)}
-J\sum_{\langle xy\rangle}(L^x_{(x)}L^x_{(y)}+L^y_{(x)}L^y_{(y)}),
\end{eqnarray}
where the coefficients are given by $U=2(U_0-W)$, 
$\widetilde{\mu}=-(\Delta_a-V_a)+(\Delta_b-V_b)$, and $J=\sqrt{V_aV_b}$.

This two-species Bose-Hubbard model can be realized in a $^{87}$Rb and $^{41}$K
Bose-Bose mixture where an inter-species Feshbach resonance is accessible.
The details of this proposal are discussed in Ref.~\cite{Zou:2014rha}.

\section{Conclusion}
In summary, we have considered the $U(1)$-Higgs model in two dimensions. 
Neglecting the plaquette interactions, we have provided an effective theory where link variables are integrated out,
producing 4-field operator. In this approximation, the Nambu-Goldstone modes have disappeared but can be reintroduced 
at first order in the plaquette interactions. Our goal is to provide a proof of principle that some 
approximate ``analog computer'' for the  $U(1)$-Higgs model can be built using cold atoms trapped in an optical lattice. 
As a first step in this direction, we discussed a recent proposal to implement the $O(2)$ model (describing the  Nambu-Goldstone modes without gauge fields) on optical lattices using a $^{87}$Rb and $^{41}$K Bose-Bose mixture of cold atoms.

\acknowledgments 
Y. M. thanks Boris Svistunov and other participants of the KITPC workshop ``Precision Many-body Physics of Strongly correlated Quantum Matter'' for valuable conversations. 
This work was supported in part by a DoD contract
Award Number W911NF-13-1-0119 and NSF grant DMR-1411345.

\end{document}